\documentclass[%
 reprint,
 amsmath,amssymb,
 aps,
]{revtex4-2}

\usepackage{graphicx}
\usepackage{dcolumn}
\usepackage{bm}
\usepackage[percent]{overpic}
\usepackage{siunitx}%
\usepackage[figuresright]{rotating}%
\usepackage{adjustbox}
\usepackage{multibib}
\usepackage{color}
\usepackage{setspace}
\usepackage{xfp}
\usepackage{booktabs}%

\newcommand\modalforce{F_{y,b}}
\newcommand\ii{\text{i}}
\newcommand\ee{\text{e}}
\newcommand\Gmean{$G^*=6.68157\times 10^{-11}\,$m$^3$kg$^{-1}$s$^{-2}$}\newcommand\Gfinal{$G^*=6.68(10)\times 10^{-11}\,$m$^3$kg$^{-1}$s$^{-2}$}\newcommand\Gstd{0.02\%}\newcommand\GSU{$1.46$\%}\newcommand\Gvar{1.32\%}\newcommand\RelUexp{$\pm2.85$\%}\newcommand\Nmeas{28 }\newcommand\Qmean{44819(535)}\newcommand\Ediss{3.75$\times 10^{-18}$~W}\newcommand\GSUsingle{$1.46$\%}\newcommand\Dev{0.1\% higher } \newcommand\crosstalkpercent{$1.00$~}\newcommand\beamdimensions{$0.01$~mm}\newcommand\rotationCenterOffset{$0.20$~mm}\newcommand\eccentricity{$1.00$~\textmu m}\newcommand\anglemisalignment{$0.03$~$^\circ$}\newcommand\masses{$100.00$~mg}\newcommand\barangledifference{$0.10$~$^\circ$}

\begin{document}

\preprint{APS/123-QED}

\title{Dynamic gravitational excitation of structural resonances in the hertz regime using two rotating bars}
\author{Tobias Brack} 
\altaffiliation{These authors contributed equally to this work.}
\author{Jonas Fankhauser}
\altaffiliation{These authors contributed equally to this work.}
\author{Bernhard Zybach}
 \altaffiliation{These authors contributed equally to this work.}

 \author{Stephan Kaufmann}
 \author{Francesco Palmegiano}
 \author{Jean-Claude Tomasina}
 \author{Stefan Blunier}
 \author{Donat Scheiwiller}
\author{Jürg Dual}%
 \email{dual@imes.mavt.ethz.ch}
\affiliation{%
 Institute for Mechanical Systems, ETH Zürich, Tannenstrasse 3, Zurich, 8092, Switzerland.
}

\author{Fadoua Balabdaoui}
\affiliation{Seminar of Statistics, ETH Zürich, Rämistrasse 101, Zurich, 8092, Switzerland.}%

\date{December 21, 2022}

\begin{abstract}
With the planning of new ambitious gravitational wave (GW) observatories, fully controlled laboratory experiments on dynamic gravitation become more and more important. Such new experiments can provide new insights in potential dynamic effects such as gravitational shielding or energy flow and might contribute to bringing light into the mystery still surrounding gravity. Here we present a laboratory-based transmitter-detector experiment using two rotating bars as transmitter and a 42\,Hz, high-Q bending beam resonator as detector. Using a highly precise phase control to synchronize the rotating bars, a dynamic gravitational field emerges that excites the bending motion with amplitudes up to 100\,nm/s or 370\,pm, which is a factor of 500 above the thermal noise. The two-transmitter design enables the investigation of different setup configurations. The detector movement is measured optically, using three commercial interferometers. Acoustical, mechanical, and electrical isolation, a temperature-stable environment, and lock-in detection are central elements of the setup. The moving load response of the detector is numerically calculated based on Newton’s law of gravitation via discrete volume integration, showing excellent agreement between measurement and theory both in amplitude and phase. The near field gravitational energy transfer is 10$^{25}$ times higher than what is expected from GW analysis.
\end{abstract}

\maketitle

\section{Introduction}\label{sec:main}
With increasing research efforts in gravitational waves and yet unexplained differences in the measurement of $G$ by different research groups, laboratory-based, dynamical gravitational experiments become more and more important. However, experiments with frequencies $> 0.1$~Hz are very rare when it comes to a full quantitative comparison between theory and experiment, such as the gravitationally induced vibration amplitude and phase of a detector system, the distance behavior, or energy flow. This is illustrated in Table~\ref{tab:OVtable} showing an overview of the few dynamic gravitational experiments reported during the last 50 years.\\
To fill this gap, we presented in a previous work the dynamical gravitational coupling between two parallel beams vibrating in their first bending resonance at 42~Hz \cite{Brack2022}. This setup allowed for a quantitative investigation of the distance behavior and the estimation of the gravitational constant $G$. Although this setup allowed for important advances in dynamic gravitation measurements, it features some drawbacks, such as the need for matching transmitter and detector resonance frequency or the generation of very low detector amplitudes.\\%
Therefore, motivated by the pioneering works of Sinsky and Weber \cite{Sinsky1967,Sinsky1968}, Astone \cite{Astone1991,Astone1998}, and the Hirakawa group \cite{Hirakawa1980,Ogawa1982,Kuroda1985}, we introduce here a novel setup which combines the experimental and analytical advantages of our initial setup with the benefits of a rotational excitation and its elaborated theory. As illustrated in Table~\ref{tab:OVtable}, all recent works use one single transmitter. In contrast, here we use \textit{two} slender, 0.5~m long tungsten bars arranged symmetrically to a bending beam resonant detector. This double transmitter arrangement allows for various configurations of the excitation, that is, rotation direction and frequency, within the same experimental setup. \\
Previous dynamic experiments encountered a relatively large measurement uncertainty for the estimation of $G$. This was mainly caused by electrical or mechanical, non-gravitational crosstalk and the fact that the measurement accuracy reaches its limit with the resulting small displacement amplitudes. In the setup presented here, some of these issues can be addressed using a precisely defined rigid body transmitter movement and considerably larger detector amplitudes. These advantages, however, come at the expense of high requirements on the balancing of the rotors and their motion control, in particular their phase accuracy and phase jitter. To demonstrate the potential of this new setup, we use the same detector beam as in \cite{Brack2022} and the same measurement procedure. Thus we obtain the detector resonance amplitude and phase at different distances, as well as an estimation of the gravitational constant $G$.\\
While the motion of the rotating bars is trivial to model, its interaction with the detector beam is categorized as a nonstandard moving load problem \cite{ Fryba1973} with varying speed and shape of the moving force pulse. Its solution is greatly simplified by the periodicity of the loading and the restriction to the steady state. Considering the length scales of the experiment, the detector is placed in the near-field of the transmitter. Since the excitation is not necessarily purely quadrupole, Newton’s law has been directly applied to all relevant mass elements. We compute the results using a highly accurate 3D finite element model and a simplified 1D model for comparison.\\ 
This article presents a theoretical and experimental description of the experiment and shows its feasibility and enormous potential as well as technical pitfalls, without providing highly accurate $G$ measurements yet. We show that the double transmitter approach provides a large variation of dynamic gravitational fields, and that the setup is able to accurately measure the resulting detector movement in amplitude and phase.
\begin{figure}[t!]
    \centering
				\includegraphics[width=0.45\textwidth]{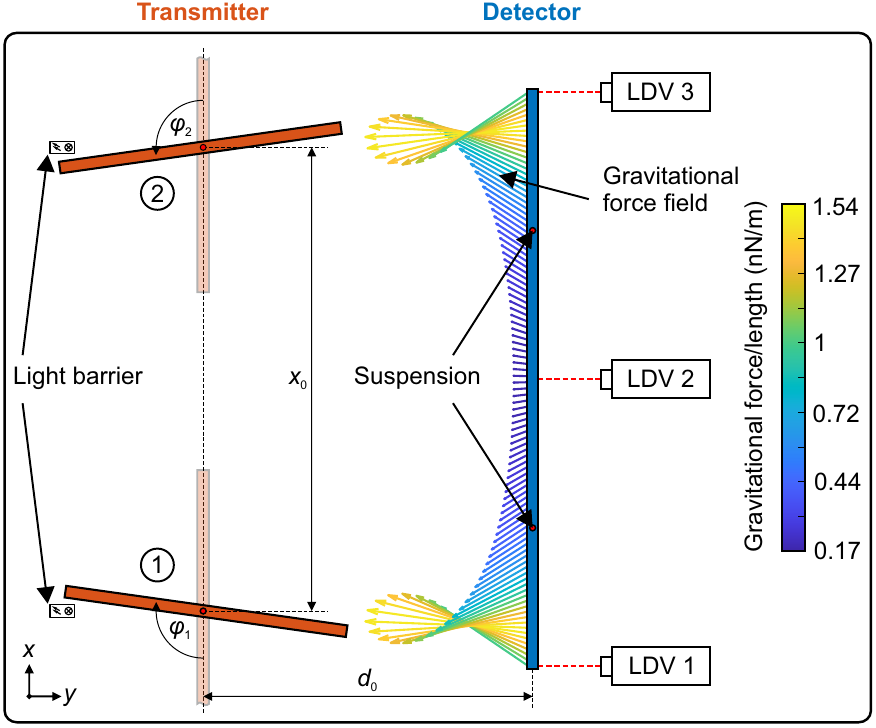}
    \caption{\textbf{Illustration of the measurement principle}. Sketch of the measurement setup using two rotating bars (orange) and one detector beam (blue), suspended in the nodal points of the first bending mode. The rotating bars create a dynamic gravitational force field on the detector that is illustrated by the colored arrows (numerical solution for $d_0= 300$\,mm, $\varphi_{1,2} = 98^\circ$). 
The vibration of the detector is measured optically by three laser Doppler vibrometers (LDV). The colorbar illustrates the gravitational force density of the dynamic force field in the $xy$-plane in nN/m.}%
    \label{fig:1}%
\end{figure}%
\begin{figure*}[t!]
  \centering   
  \begin{overpic}[width=0.9\textwidth]{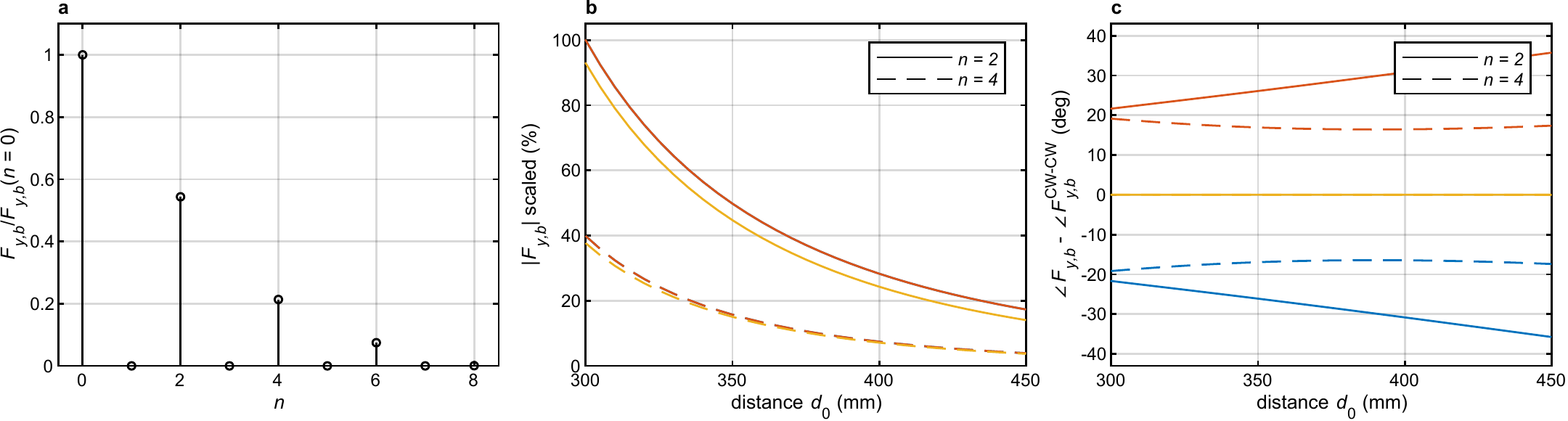}
     \put(55,13){\includegraphics[scale=0.5]{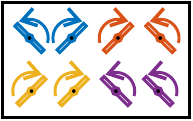}}  
	   \put(88,11){\includegraphics[scale=0.5]{legend}}
  \end{overpic}
\caption{\textbf{Properties of the gravitationally induced modal force} $\modalforce$ acting on the detector beam, produced by two identical rotating bars. Analytical results, 1D approximation. \textbf{a}, Scaled amplitude spectrum at multiples $n$ of the rotation frequency, $\Omega$, exemplarily for opposite rotation direction and $d_0=300$\,mm. \textbf{b/c}, Second (solid line) and fourth (dashed line) harmonic as a function of transmitter/detector distance, $d_0$, for various setup configurations. \textbf{b}, Amplitude scaled to the highest value at $n=2$, opposite rotation, $d_0=300$\,mm. Note: some lines are hidden due to equal results for same (yellow, purple) and opposite (red, blue) rotation direction. \textbf{c}, Phase relative to the phase of same rotation direction (yellow, purple).}
\label{fig:2}%
\end{figure*}
\section{Theory of dynamic gravitational force fields generated by rotating bar}\label{sec:theory}
The setup presented in this article consists of a transmitter and detector system, cf. Fig \ref{fig:1}a. The transmitter system is composed of two identical, slender bars of square cross section that can rotate around the $z$-axis in their center of mass. The bars are arranged symmetrically to the middle of the detector beam at a distance $d_0$, that is, the $y$-distance between the central axis of the detector and the centers of rotation. The detector is designed as a resonator, with the first lateral bending mode to be the relevant vibration mode. We selected a slender beam of rectangular cross section and twice the length of the transmitter bars. Due to the rectangular shape, it is ensured that horizontal and vertical bending do not have the same resonance frequency.\\
When the bars rotate, a dynamic gravitational force field develops due to the periodically changing distance between detector and transmitter mass elements. Hence, given that a suitable rotation frequency is selected, the dynamic gravitational force field periodically excites the first bending mode of the detector beam, which then develops a measurable deflection.\\
To investigate the force field acting on the detector, Newton's law of gravitation is applied to each pair of transmitter/detector mass elements. The transmitter rotation is maintained at a constant frequency with very high accuracy using active rotation control (cf. supp. material). The much smaller gravitational forces acting on the transmitter bars do not affect the rotation. Newton's law of gravitation yields a time dependent, three-dimensional gravitational body force on the detector generated by each bar. For a first analytical analysis of the gravitational effects a 1D approach is sought, where the body forces in $y$-direction are approximated by two line forces $f_{y}^{(k)}(x_d)$ acting on the central axis of the slender detector beam, where $x_d$ denotes the $x$ coordinate of the detector beam center line and $k$ the number of the rotating bar, cf. Fig.~\ref{fig:1}. The detector's bending movement can then be calculated using an Euler-Bernoulli beam model and a modal approach \cite{hagedorn2007,Brack2022}, in which the distributed force is reduced to the effective modal excitation force $\modalforce$ of the first bending mode via
\begin{eqnarray}
\modalforce=\int_0^{l_d}\left(f_{y}^{(1)}(x_d)+f_{y}^{(2)}(x_d)\right)U_{b,1}(x_d)\text{d}x_d\, ,
\label{eq:one}
\end{eqnarray}
where $U_{b,1}$is the mode shape function of the first bending mode, normalized so that $U_{b,1}(0) = 1$. \\
To assess the frequency components of the excitation, a Fourier series is computed. Due to the nonlinear law of gravitation, higher harmonics must occur, which is illustrated by Fig.~\ref{fig:2}a, where the spectral components at multiples of the rotation frequency $\Omega$ are shown qualitatively (scaled to the DC value). It can be observed that the force field contains frequency components at even multiples of the rotation frequency only with decreasing contribution. Hence the excitation frequency $\Omega$ must be chosen 1/2 or 1/4 of the detector beam's resonance frequency $\omega_0$ to achieve maximum detector excitation.\\
The use of higher harmonics as excitation and the possibility of different rotation configurations (one/two bars, same/opposite direction, pos./neg. direction) enables different setup configurations that will be discussed briefly. Due to the rotation, the force distribution at the detector beam varies in time and space and can be categorized as a nonstandard periodic moving load problem \cite{Fryba1973}. Considering one rotating bar only, the variation of the force distribution can be illustrated by a transverse pulse with varying speed and shape moving along the $x$-axis of the detector. Considering two bars rotating in opposite directions, the situation becomes symmetric and the net force in $x$-direction becomes zero. While the strength, that is, the amplitude of the force field, decreases with increasing transmitter/detector distance $d_0$, the behavior of the phase between transmitter rotation and detector vibration is less intuitive. Interpreting the phase as a measure for the angle of maximal excitation, it becomes obvious that said angle must not necessarily be at $\varphi_k = 90^\circ$ due to the mode shape of the first bending mode and that it will change with distance $d_0$.\\
\begin{figure*}[th!]
    \centering
		\includegraphics[width=0.9\textwidth]{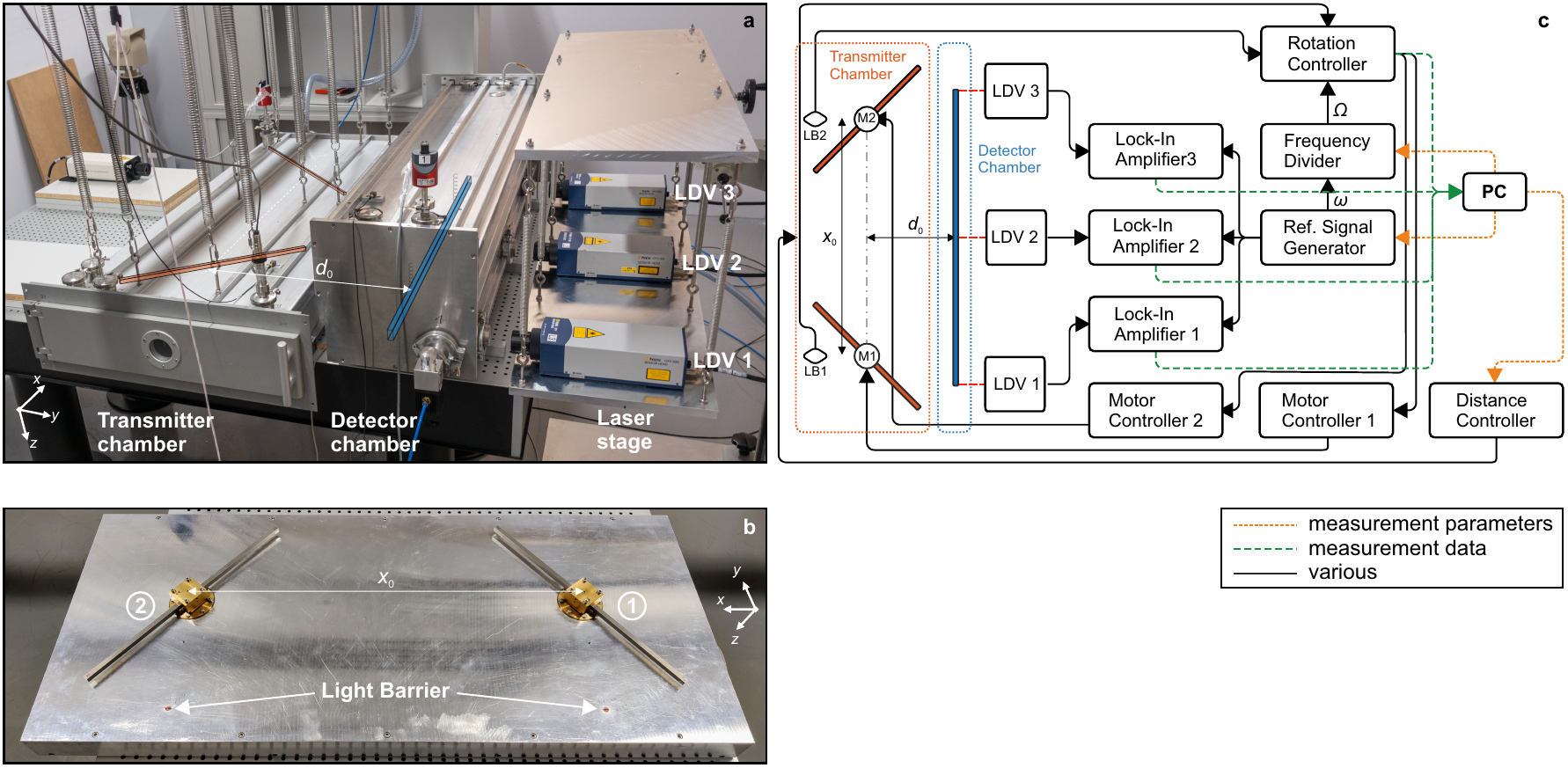} 
    \caption{\textbf{Illustration of the measurement setup.} \textbf{a}, Photo of the setup: The rotating transmitter bars (red drawing) are located inside the transmitter chamber. The chamber itself is hanging from springs attached to a carrier bar movable in $y$-direction. Inside the detector chamber the detector beam (blue) is hanging on rubber wires attached at the nodal points of the first bending mode. The detector chamber is fixed to an anti-vibration table, isolating the beam from the environment, and minimizing transmission of non-gravitational forces. The detector movement is measured using three laser Doppler vibrometers (LDV), positioned on a separate stage, likewise isolated via springs. The distance $d_0$ is varied by moving the transmitter chamber. The whole setup is located in an underground laboratory, providing excellent temperature stability and minimal seismic noise. \textbf{b}, Separate view of the rotating bars mount with the light barriers used for the rotation control. \textbf{c}, Block diagram of the operation scheme: orange lines illustrate control signals from the PC that set the distance $d_0$, excitation frequency $\Omega$ and the frequency ratio $\omega/\Omega$. Green lines mark measurement signals.}%
    \label{fig:3}%
\end{figure*}
When both bars rotate in the same direction, the maximum of the two force fields do not occur at the same time, therefore the force fields don't superpose symmetrically, resulting in an approx. 8\% smaller amplitude and a non-zero net force in $x$-direction. For the same reason, the angle of maximal excitation occurs exactly at $\varphi_k =90^\circ$, which eliminates the distance dependency of the phase. A phasor representation of the superposition of the modal forces is presented in the supplementary material.\\
Figs.~\ref{fig:2}b/c show the result of an analytical calculation of $\modalforce$ as a function of distance, illustrating the aforementioned effects. As it was expected from the spectrum in Fig.~\ref{fig:2}a, an excitation with $\Omega = \omega_0/4$ creates an amplitude of about 40\% lower compared to $\Omega = \omega_0/2$.\\
To numerically assess the detector movement caused by the gravitational field, the modal force is applied as excitation force to the well-known equation of motion of a free–free Euler–Bernoulli beam. It can be shown that the complex detector bending velocity at resonance, $v_{b,0}$ can be described via
\begin{equation}
v_{b,0} = \frac{GQ_d}{\omega_0}\Gamma\approx \frac{GQ_d}{\omega_0d_0^a}\gamma\, ,
\label{eq:2}
\end{equation}
where $G$ is the gravitational constant, $Q_d$ the Q factor of the detector beam and $\omega_0$ its first bending resonance frequency. $\Gamma$ is a complex numerical coefficient which depends on the distance $d_0$, the detector beam and rotating bar's dimensions and their relative position $x_0$, the rotating bar's masses and mass distributions, the ratio $n=\omega_0/\Omega$, and the rotation configuration. It is calculated numerically for any given configuration. To avoid errors due to the 1D approximation, the calculation of $\Gamma$ is based on a full 3D model (cf. supp. material), for which Eq.~\ref{eq:2} holds as well.\\
The modulus of $\Gamma$ can be approximated to follow a power law distance behavior $\Gamma\approx\gamma/d_0^a$. It has to be noted that the power law coefficient $a$ considerably differs from the prediction $a= 5$ of the dynamic gravitational field around a rotating bar detected by a resonant antenna (quadrupole-quadrupole interaction) \cite{Oide1980,Hirakawa1976}.
The values of $a$ for the setup, length scales, and configurations presented in this article are summarized in Table~\ref{tab:2}.\\
The gravitational constant $G$ can be estimated from Eq.~\ref{eq:2}, using the measured detector amplitude $v_{b,0}$, the detector's vibration properties ($Q_d,\,\omega_0$), and the numerical coefficient $\Gamma$. Since this will result in a complex number, only the modulus is taken as estimation for $G$. The angle gives information about the phase difference between measurement and theory, cf. Fig.~\ref{sfig:2}.
\section{\label{sec:setup}Experiment design and measurement procedure}
\begin{figure*}[th!]
    \centering
		  \begin{overpic}[width=0.9\textwidth]{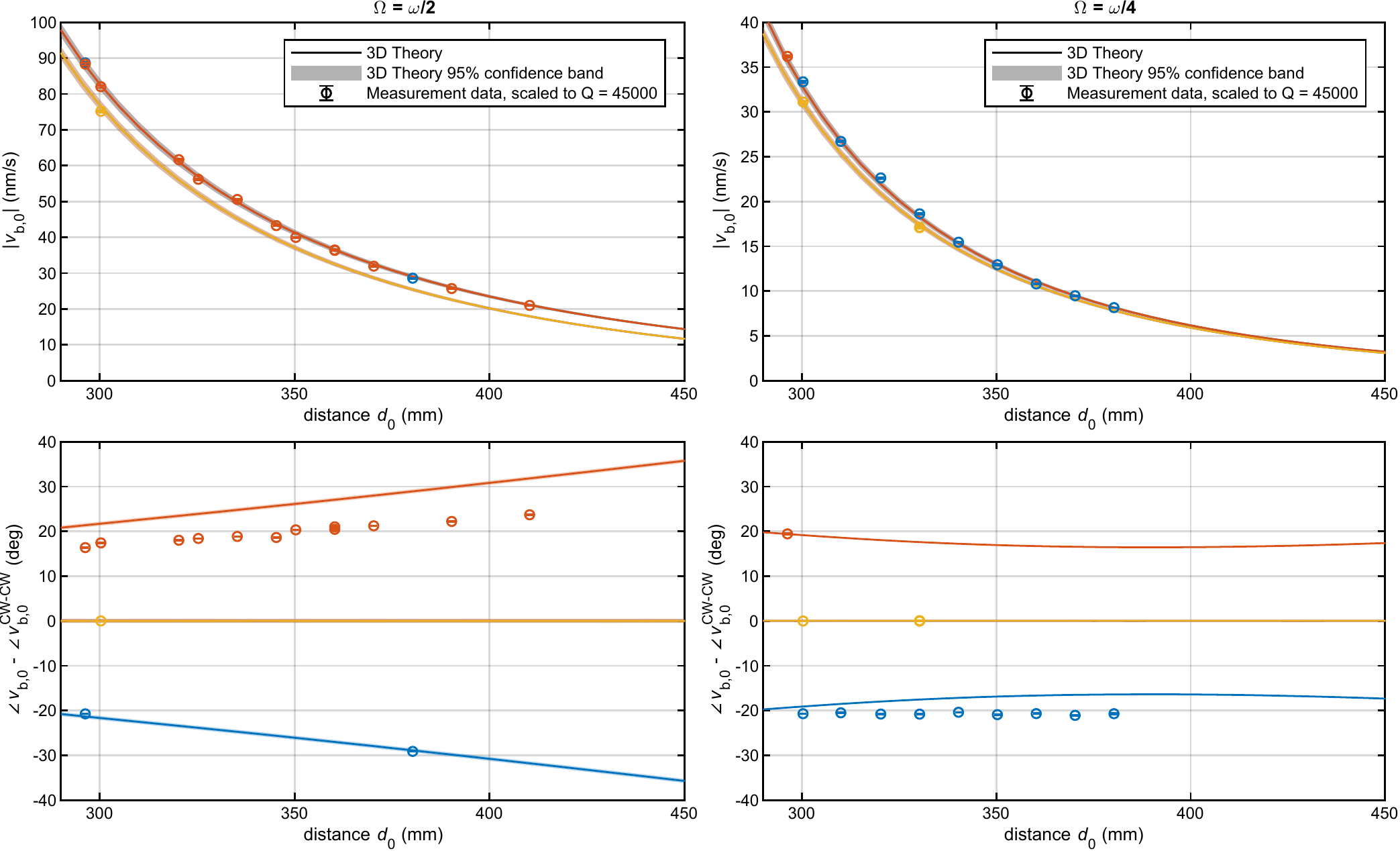}
     \put(38.4,46){\includegraphics[scale=0.5]{legend}} 
		\put(88.4,46){\includegraphics[scale=0.5]{legend}} 
		\put(38.4,17){\includegraphics[scale=0.5]{legend}} 
			\put(88.4,22){\includegraphics[scale=0.5]{legend}} 
  \end{overpic}
    \caption{\textbf{Measurement results of gravitationally induced detector beam motion}. Compilation of results at different setup configurations and comparison with the 3D theory solution. The data points represent the detector motion at resonance, obtained from a SDOF fit of a 24-point frequency response measurement around the detector's resonance. The measurement data point error bars comprise uncertainties that affect $v_{b,0}$ in Eq.~\ref{eq:2}, that is, the SDOF fit, laser vibrometer and lock-in amplifier. The 95\% confidence band of the theoretic result is derived from the standard uncertainties of the remaining parameters, $G\,,\omega_0\,,Q_d\,,\Gamma$. \textbf{Left column}, excitation with second harmonic $\Omega=\omega/2$. \textbf{Right column}, excitation with fourth harmonic $\Omega=\omega/4$. \textbf{Upper row}, Detector beam bending amplitude $\lvert v_{b,0}\rvert$ as a function of the beam distance $d_0$. A power-law behavior can be identified, the exponents are given in Table~\ref{tab:2}. The values lie mostly within the confidence band (shaded area) of the theoretical prediction. \textbf{Lower row}, Phase difference between rotation and detector response, relative to the phase for equal rotation direction (yellow/purple line).}%
    \label{fig:4}%
\end{figure*}
\begin{figure*}[t!]
    \centering
						  \begin{overpic}[width=0.9\textwidth]{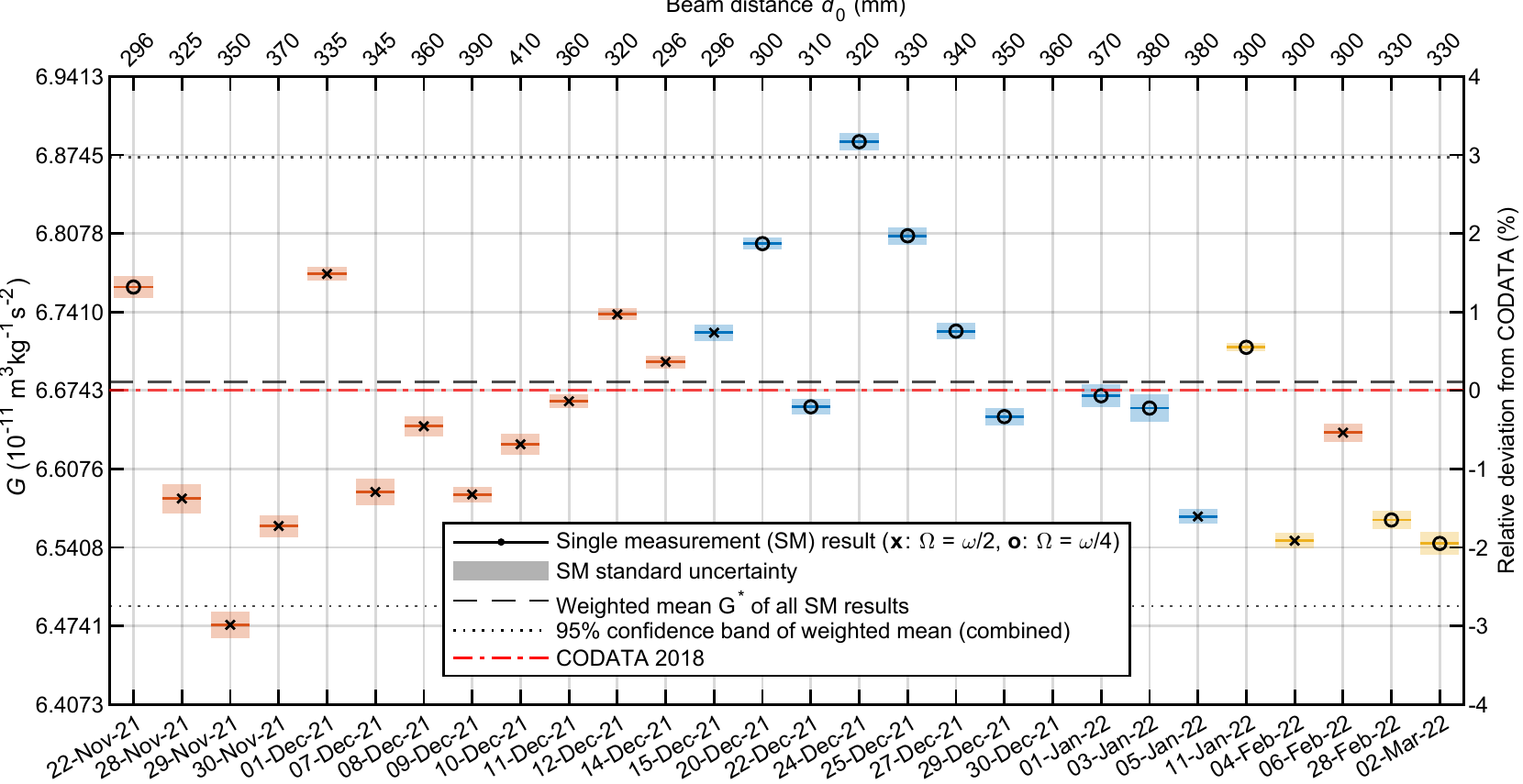}
			\put(85.8,40){\includegraphics[scale=0.5]{legend}}
			  \end{overpic}
    \caption{\textbf{Measurement result of the gravitational constant.} Over a period of about 2 months, \Nmeas~single measurement runs were conducted at different beam distances $d_0$ and setup configurations. The rotation configuration is color coded, while the marker represent excitation with the second (\textbf{x}) or fourth (\textbf{o}) harmonic. Each measurement yields the resonance amplitude and phase of the detector, whereas only the amplitude has been used to estimate the gravitational constant using the analytical 3D model, cf. Eq.~\ref{eq:2}. The figure illustrates the single and averaged results for $G$ (left y axis) and the deviation from the CODATA 2018 value (right y axis). Lower x axis, measurement date; upper x axis, beam distance $d_0$. An inverse-variance weighted mean (black dashed line)
		yields an overall estimate of \Gmean, with a combined standard uncertainty of about \GSU. The estimation is about \Dev than the CODATA 2018 value (red dash–dot line). The plotted 95\% confidence band (black dotted lines) represents the extended combined measurement uncertainty ($k = 1.96$) based on statistical and systematic uncertainties (Table~\ref{tab:1}).}%
    \label{fig:5}%
\end{figure*}
\begin{table*}[ht!]
	\centering
\begin{tabular}{lll}

																			&\textbf{Standard uncertainty } &  $\mathbf{\Delta G/G}$ \textbf{(\%)} \\
\hline 
\textbf{Systematic errors} & & \\Laser vibrometer amplitude & 0.67 \% & 0.39 \% \\Laser angular misalignment & 0.01 \% & 0.01 \% \\Lock-in amplifier & 0.01 \% & 0.00 \% \\Transmitter/detector distance, $d_0$ & 0.17 \% & 0.98 \% \\Mechanical crosstalk & 1.00 \% & 1.00 \% \\Model parameter amplitude, $\lvert\gamma\rvert$ & 0.14 \% & 0.14 \% \\Model parameter phase, $\angle(\gamma)$ & 0.24 deg & - \\Transmitter/detector dimensions & 0.01 mm & -$^*$ \\Transmitter rotation centre $x$/$z$ offset & 0.20 mm & -$^*$ \\Transmitter bar eccentricity & 1.00 \textmu m & -$^*$ \\Transmitter/detector angular misalignment & 0.03 $^\circ$ & -$^*$ \\Transmitter/detector mass & 100.00 mg & -$^*$ \\Transmitter bar angle difference & 0.10 $^\circ$ & -$^*$ \\\textbf{Statistical errors} & & \\Detector bending amplitude at resonance, $v_{b,0}$ & 0.10 \% & 0.10 \% \\Detector beam resonance frequency, $\omega_0$ & \textless\,0.01 \% & \textless\,0.01 \% \\Detector beam phase shift at resonance & 0.11 deg & - \\Detector beam Q factor, $Q_d$ & 0.18 \% & 0.18 \% \\Transmitter/detector distance, $d_0$ & \textless\,0.01 \% & \textless\,0.01 \% \\Statistical errors of all measurements & & 0.02 \% \\\textbf{Combined error} & \textbf{Single measurement} & \textbf{1.46 \%} \\                         & \textbf{All measurements} & \textbf{1.46 \%} 
\end{tabular}
    \caption{One-sigma error budget used for the assessment of the combined measurement uncertainty of $G$ estimated from one single measurement and from all \Nmeas measurements. For parameters where different values apply, e.g. distance errors, the highest value of the uncertainty is reported. Note that the phase uncertainties do not contribute to $\Delta G$.\\
		$^*$Included in uncertainty of model parameter $\gamma$.}%
    \label{tab:1}%
\end{table*}		
Figure~\ref{fig:3}a shows the experimental setup as it was already introduced and used in \cite{Brack2022}, now with the rotating bars located in the transmitter chamber. The transmitter bars are made of tungsten both with length of 499.86(1)\,mm and a quadratic cross section of 10.06(1)\,mm edge length, each driven by a 50W maxon EC-i 40 brushless motor to rotate with frequency $\Omega$. The distance between the rotation centers is $x_0=800.0(3)$\,mm. The bars' mass has been measured equal to 970.8(1)\,g and 971.8(1)\,g, respectively. The detector is made of titanium with dimensions 1000.00(1)\,mm\,\texttimes\,16.97(1)\,mm\,\texttimes\,8.49(1)\,mm and a Q factor of $Q_d=\Qmean$ at a detector chamber pressure of about 0.03\,mbar. The detector is hanging on two EPDM rubber strings glued to the beam at the nodal points of the first bending mode, which minimizes both external damping and force transmission, and enables the use of a free-free beam model. A mass of 7.1\,g at the center of each rubber wire provides additional decoupling. The beam's movement is measured optically using three Polytec OFV-5000/505 laser Doppler interferometers (LDV) placed on a separate, spring-suspended platform. The output signals of the laser interferometers are fed into individual digital lock-in amplifiers (Zurich Instruments MFLI) via a 12\,dB attenuator. The lock-in amplifiers use eight cascaded low-pass filters with time constant of 31\,s to extract the velocity amplitude and phase at the measurement frequency $\omega$. The reference is provided by a high precision signal generator (SRS FS740 with Rubidium time base, frequency error $< 10$\,pHz, phase accuracy $<1$\,ns). From the movement at three different positions one can distinguish between two rigid body motions (translation in $y$ and rotation around $z$) and the lateral bending motion, assuming a known bending mode shape $U_{b,1}$. \\
Both detector and transmitter system have been placed in separate aluminum vacuum chambers to avoid acoustic coupling effects or any excitation other than gravitation. Using two separate Edwards nXDS 10i vacuum scroll pumps running continuously, the pressure inside the detector and transmitter chambers was kept constant at 0.027(3)\,mbar and 0.103(5)\,mbar, respectively. No disturbing influence of the vibrations produced by the pumps was identified. The chamber containing the detector was fixed to a vibration isolation table using an 80\,mm thick aluminum base plate of 70\,kg mass. The transmitter chamber is hanging on steel springs (Durovis 20/8/5) from a movable bar attached to a solid frame with high damping. This way, the distance $d_0$ can be varied between 0.3\,m and 0.6\,m. Accelerometers mounted on both chambers give information about remaining movement of the chambers.\\
To investigate the gravitational coupling and to compare it with the numerical solution, the frequency response of the detector is measured around the first bending resonance $\omega_0$. We use a 24-point frequency sweep with step times of 75\,min to account for the large time constant of the detector beam ($\tau \approx 5.5$\,min) and the lock-in amplifiers (99\% settling time approx. 9\,min). By averaging the last 16\,min of each step, a signal-to-noise ratio (SNR) ratio of up to 500 can be achieved. After fitting the frequency response of a single-degree-of freedom (SDOF) oscillator \cite{Brack2022}, the following parameters are obtained: amplitude and phase at resonance, resonance frequency, the detector's Q factor, and a complex offset constant.\\%
The low bandwidth of the detector and the need for synchronization of both beams imposes extremely high requirements on the resolution and stability of the rotation frequencies. Since this cannot be realized using the motor's built-in encoder, a custom rotation control and synchronization based on two light barriers (LB) have been implemented, cf. Fig.~\ref{fig:3}b/c. The rotation control provides a stable rotation with a RMS phase jitter $<0.03^\circ$. Details on the rotation control can be found in the supplementary material.\\
Due to the long-time measurement, it must be ensured that resonance frequency and Q factor of the detector beam remain stable during the measurement period. Therefore, the whole setup has been placed in an underground laboratory in the Swiss Alps where a very stable temperature can be guaranteed, resulting in an average temperature span of 0.004\,$^\circ$C per measurement point (75\,min). Remaining temperature variations have been subsequently compensated based on the linear temperature dependency of the frequency. Further details can be found in \cite{Brack2022}.\\
Despite rotating with a fraction of the measurement frequency $\omega$, minimal unbalances of the bars, motor friction, material inhomogeneities etc. can generate small vibrations of the transmitter chamber with amplitudes up to 2 \textmu m/s (8\,nm) at $\omega$. These disturbances can propagate to the detector chamber most likely via acoustical transmission and structure borne sound. Therefore, as the decoupling of the detector is not perfect, unwanted, non-gravitationally induced detector vibration can occur. Accelerometers mounted on both chambers reveal that the use of $\Omega = \omega/2$ produces detector chamber velocity amplitudes of approx. 2\,nm/s at the frequency of measurement, while the detector chamber moves with amplitudes less than 0.5\,nm/s at $\omega$ when rotating with $\Omega = \omega/4$, cf. Fig.~\ref{sfig:1}.

\begin{table*}[th!]
	\centering
\begin{tabular}{lccccc}
& \textbf{CCW-CW}& \textbf{CW-CCW}& \textbf{CCW-CCW}& \textbf{CW-CW}& \textbf{single bar*}\\ \hline$\Omega = \omega/2$& 4.35& 4.35& 4.65& 4.65& 4.35\\ $\Omega = \omega/4$& 5.83& 5.83& 5.78& 5.78& 5.83
\end{tabular}
    \caption{Value of the power law exponent $a$ as introduced in Eq.~\ref{eq:2} for distances $d_0$ between 300 and 600 mm. Coefficients obtained from a power law fit based on 3D numerical results. The fit's average coefficient of determination is $R^2 = 99.97\%$ (34 values). CW = clockwise; CCW = counterclockwise.\\
		$^*$Equal values for all single bar variations.}%
    \label{tab:2}%
\end{table*}
\subsection{\label{sec:results}Examination of gravitationally induced detector vibration}
Over a period of two months, \Nmeas measurement runs were conducted at different beam distances and setup configurations. Each measurement corresponds to a 24-point frequency sweep of 36\,h duration, where resonance amplitude and phase, as well as the resonance frequency and Q-factor, have been extracted by fitting the frequency response function of a SDOF oscillator. To detect measurements that were affected by unstable conditions, the following quality criteria have been applied: Detector chamber pressure span $< 0.01$\,mbar, detector chamber temperature span $< 0.1^\circ$C, SDOF fit coefficient of determination $R^2>99\%$.\\
The results of the measured complex bending velocity at resonance $v_{b,0}$ are summarized in Fig.~\ref{fig:4}. Additionally, the results of the 3D simulation are displayed. Since the simulation uses a fixed detector Q factor and resonance frequency, the measured amplitudes are scaled to these values using the linear dependency of the velocity on $Q$ and $\omega_0^{-1}$, cf. Eq.~\ref{eq:2}. As shown in the first row of Fig.~\ref{fig:4}, the measured amplitudes match the numerical prediction very well for all setup configurations. Considering the uncertainty of the theoretical prediction that comprises statistical and systematic uncertainties as summarized in Table~\ref{tab:1}, the measured amplitudes mostly lie within the 95\% tolerance band of the theory/simulation. In the second row of Fig.~\ref{fig:4}, experimental and theoretical results of the phase of the detector vibration are depicted. Since the phase for configurations using the same rotation direction (yellow lines) should be independent on the distance, it has been used as reference. The experimental phase values show the predicted number range and trend, but do not match the theory exactly. 
Besides the measurement system that might introduce some uncertainties, mechanical crosstalk, which is more pronounced in the phase, is suspected to be the main reason for the observed deviation. A small movement of the detector chamber was measured, as shown in Fig.~\ref{sfig:1}, where the average velocity amplitudes of both detector and transmitter chamber are displayed. However, the mechanism and magnitude of the force transmission from detector chamber to detector beam are not yet fully understood. Therefore, a rough estimate of a 1\% contribution to the systematic uncertainties due to mechanical crosstalk has been added to the error budget.\\%
Finally, $G$ was estimated from each measurement result using Eq.~\ref{eq:2} (modulus), where the detector's Q factor has been incorporated individually for each measurement. In Fig.~\ref{fig:5}, the single results are depicted as mean value and standard deviation from statistical errors (colored patches). Combining the single measurements of $G$ by means of inverse-variance weighting yields \Gmean~with a relative standard deviation of \Gstd~(black dashed line in Fig.~\ref{fig:5}). The overall 95\% confidence band (black dotted line in Fig.~\ref{fig:5}) represents the extended combined measurement uncertainty (k = 1.96) based on the statistical and systematic uncertainties as summarized in Table~\ref{tab:1}. Finally, $G$ is estimated with \Gfinal. We would like to note that although the mean value we obtain is only about \Dev than the CODATA 2018 value \cite{Tiesinga2021}, the single values obtained from the measurements have a standard deviation which is \Gvar~of the mean. This can be mostly attributed to the uncertainty of the measurement chain which is relatively high at this stage of the project (95\% confidence band = \RelUexp). Figure~\ref{sfig:2} illustrates the phase difference between measurement and theory, increasing with larger distances, supporting the assumption of remaining mechanical, non-gravitational coupling.\\%
Based on a power balance analysis, the near field gravitational energy flow between transmitter and detector can be computed \cite{Brack2022}. At steady state, incoming energy at the detector is dissipated according to the Q factor of the detector. If all this energy is attributed to gravitation, this yields a gravitational power of maximal \Ediss~at $d_0=296$\,mm, $\Omega = \omega_0/2$, CW-CCW rotation. This is about $10^{25}$ times higher than the expected power of gravitational waves radiated from two equivalent quadrupole gravitational wave generators \cite{misner2017}. 
		\section{\label{sec:outlook}Discussion and outlook}
In the last couple of years, fully characterized dynamic gravitation experiments turned out to be a promising approach to better understand gravitational interactions \cite{Rothleitner2022}. They open up the path to an investigation of dynamic gravitational effects in a frequency range $> 1$~Hz, covering for example the gravitational wave high-frequency band \cite{Thorne95}.\\
The combination of two similar rotating bars as transmitter system and a bending beam as detector system proves to be a considerable improvement to previous experiments, cf. Table~\ref{tab:OVtable}. Besides considerably higher amplitudes of one order of magnitude, the double-transmitter setup enables the investigation of numerous setup configurations, based on rotation direction combinations and the use of different harmonics as excitation. Therefore, gravitational and remaining non-gravitational coupling can be better distinguished, making the results more reliable. The extremely precise rotation control, necessary for the double excitation, has the additional merit of enabling a precise phase measurement between rotation and detector response, which can be very helpful to further understand dynamic, non-gravitational coupling. The work presented here establishes a new experiment, however, it does not yet claim to be highly precise. Nonetheless, the observed discrepancy between theory and measurement is considerably less than 1\%.\\
To further increase the measurement quality and reliability, future work focuses on improvements that comprise a more precise distance control, reduced temperature sensitivity, characterization of mass distributions, passive and active crosstalk cancellation, and improved model accuracy. The latter requires narrower tolerances and a better understanding of the material structure and behavior of both the rotating bars (material homogeneity, ultra-precise mass measurement, etc.) and detector (temperature behavior, influences on structural damping, etc.). Since the setup allows for the use of even larger transmitter bars the signal level can be increased by about one order of magnitude. Although higher amplitudes reduce problems associated with the optical vibration measurement of such small amplitudes \cite{Brack2022,Hou1992,Weichert2012,Pisani2012}, an individual calibration of the interferometric measurement system in the nm/s range is still necessary to considerably reduce the measurement uncertainty. Alternatively, a custom made demodulation of the laser output might enable to directly trace back the detector displacement to the wavelength of light. \\
We believe that these improvements can bring the experiment way beyond a proof-of-concept state towards a highly precise measurement that might become the new standard dynamic gravitational experiment. It will help to reveal new insights in dynamic gravitation, such as frequency dependency or amplitude/phase effects due to objects between transmitter and detector (gravitational shielding). The results and findings of this and future related works can also help to advance the research and application of Newtonian calibrators that are used in gravitational-wave detectors \cite{Ross2021,Kawasaki2020,Matone2007,Astone1998,Mio1987,Hirakawa1980}.
		\clearpage
		\newpage
\def\arraystretch{1.1}%
\begin{sidewaystable}
\centering
\resizebox{\textwidth}{!}{%
\footnotesize
\setlength{\extrarowheight}{0.1em}
\begin{tabular}{c|ccc|cccccc|ccc|cc}
\hline
\begin{tabular}[c]{@{}c@{}}\phantom{a}\\\phantom{a} \end{tabular} & 
  \multicolumn{3}{c|}{Transmitter properties} &
  \multicolumn{6}{c|}{Receiver properties} &
  \multicolumn{3}{c|}{Measurement parameters} &
  \multicolumn{2}{c}{Evaluation} \\ \hline
Work &
  Design &
  Mode &
  \begin{tabular}[c]{@{}c@{}}Mass \\ (kg)\end{tabular} &
  Design &
  Mode &
  \begin{tabular}[c]{@{}c@{}}Sensor\\ principle\end{tabular} &
  \begin{tabular}[c]{@{}c@{}}Freq.\\ (Hz)\end{tabular} &
  Q &
  \begin{tabular}[c]{@{}c@{}}Mass\\ (kg)\end{tabular} &
  \begin{tabular}[c]{@{}c@{}}TX/RX\\ distance\\ (m)\end{tabular} &
  \begin{tabular}[c]{@{}c@{}}Number of\\ force field\\ harmonics\end{tabular} &
  \begin{tabular}[c]{@{}c@{}}Number of\\ excitation\\ configurations\\ (investigated)\end{tabular} &
  \begin{tabular}[c]{@{}c@{}}Phase\\ shift\end{tabular} &
  G \\ \hline
	Sinsky68 \cite{Sinsky1968}&
  \begin{tabular}[c]{@{}c@{}}Aluminum\\ cylinder\end{tabular} &
  \begin{tabular}[c]{@{}c@{}}Longitudinal\\ resonance\end{tabular} &
  136 &
  \begin{tabular}[c]{@{}c@{}}Aluminum\\ cylinder\end{tabular} &
  \begin{tabular}[c]{@{}c@{}}Longitudinal\\ resonance\end{tabular} &
  Piezoelectric &
  1660 &
  $\approx$ 1E5 &
  1500 &
  1.72\ldots1.92 &
  1 &
  1 (1) &
  No &
  No \\
	Hirakawa80 \cite{Hirakawa1980}&
  Steel bar &
  Rotation &
  44 &
  \begin{tabular}[c]{@{}c@{}}Mass quadrupole\\ antenna\end{tabular} &
  \begin{tabular}[c]{@{}c@{}}Structural\\ resonance\end{tabular} &
  Electrostatic &
  60.5 &
  4100 &
  1400 &
  2.1\ldots4.2 &
  1 &
  2 (1) &
  No &
  No \\
	Ogawa82 \cite{Ogawa1982}&
  Steel bar &
  Rotation &
  401 &
  \begin{tabular}[c]{@{}c@{}}Mass quadrupole\\ antenna\end{tabular} &
  \begin{tabular}[c]{@{}c@{}}Structural\\ resonance\end{tabular} &
  Electrostatic &
  60.8 &
  5300 &
  1400 &
  2.6\ldots10.6 &
  1 &
  2 (1) &
  No &
  No \\
	Kuroda85 \cite{Kuroda1985}&
  \begin{tabular}[c]{@{}c@{}}Aluminum disk\\ lead-filled holes\end{tabular} &
  Rotation &
  $\approx 1$ &
  \begin{tabular}[c]{@{}c@{}}Torsional\\ antenna\end{tabular} &
  \begin{tabular}[c]{@{}c@{}}Torsional\\ resonance\end{tabular} &
  Electrostatic &
  \begin{tabular}[c]{@{}c@{}}61\\ 96\end{tabular} &
  $\approx$ 1E4 &
  \begin{tabular}[c]{@{}c@{}}0.85\\ 15\end{tabular} &
  0.1\ldots0.3 &
  1 &
  2 (1) &
  No &
  \textbf{Yes} \\
	Astone98 \cite{Astone1998}&
  \begin{tabular}[c]{@{}c@{}}Aluminum\\ constant stress \\ bar\end{tabular} &
  Rotation &
  14 &
  \begin{tabular}[c]{@{}c@{}}Cryogenic \\ aluminum \\ cylinder \cite{Astone1993}\end{tabular} &
  \begin{tabular}[c]{@{}c@{}}Longitudinal\\ resonance\end{tabular} &
  Capacitive &
  \begin{tabular}[c]{@{}c@{}}910\\ 930\end{tabular} &
  \begin{tabular}[c]{@{}c@{}}1.1E6\\ 5.6E6\end{tabular} &
  2270 &
  1.9\ldots3.5 &
  1 &
  2 (1) &
  \textbf{Yes} &
  No \\
	Ross21 \cite{Ross2021}&
  \begin{tabular}[c]{@{}c@{}}Aluminum disk\\ void/tungsten-\\ filled holes\end{tabular} &
  Rotation &
  1 &
  Test mass &
  Displacement &
  \begin{tabular}[c]{@{}c@{}}Optical (GW \\ detector \cite{Aasi2015})\end{tabular} &
  8\ldots30 &
  - &
  39.7 &
  1.18 &
  2 &
  2 (1) &
  No &
  \textbf{Yes} \\
	Brack22 \cite{Brack2022}&
  Tungsten beam &
  \begin{tabular}[c]{@{}c@{}}Bending\\ resonance\end{tabular} &
  3.9 &
  \begin{tabular}[c]{@{}c@{}}Titanium\\ beam\end{tabular} &
  \begin{tabular}[c]{@{}c@{}}Bending\\ resonance\end{tabular} &
  Optical &
  42.6 &
  3.5E4 &
  0.6 &
  0.06\ldots0.12 &
  1 &
  2 (1) &
  \textbf{Yes} &
  \textbf{Yes} \\
This work &
  Two tungsten bars &
  Rotation &
  2$\times$1 &
  \begin{tabular}[c]{@{}c@{}}Titanium\\ beam\end{tabular} &
  \begin{tabular}[c]{@{}c@{}}Bending\\ resonance\end{tabular} &
  Optical &
  42.6 &
  4.5E4 &
  0.6 &
  0.3\ldots0.42 &
  2 &
  8 (4) &
  \textbf{Yes} &
  \textbf{Yes} \\ \hline
\end{tabular}
}%
\caption{Overview of past transmitter/detector macro-scale experiments that investigate dynamic gravitational forces. Numbers are rounded.}
\label{tab:OVtable}
\end{sidewaystable}
\clearpage
\begin{acknowledgments}
We gratefully acknowledge the support of ETH Zurich, maxon motor ag, ZC Ziegler Consultants AG, and Zurich Instruments AG.
\end{acknowledgments}

\appendix

\section{Superposition of force fields}\label{sec:superforce}
\begin{figure*}[h!]
    \centering
		\includegraphics[width=0.9\textwidth]{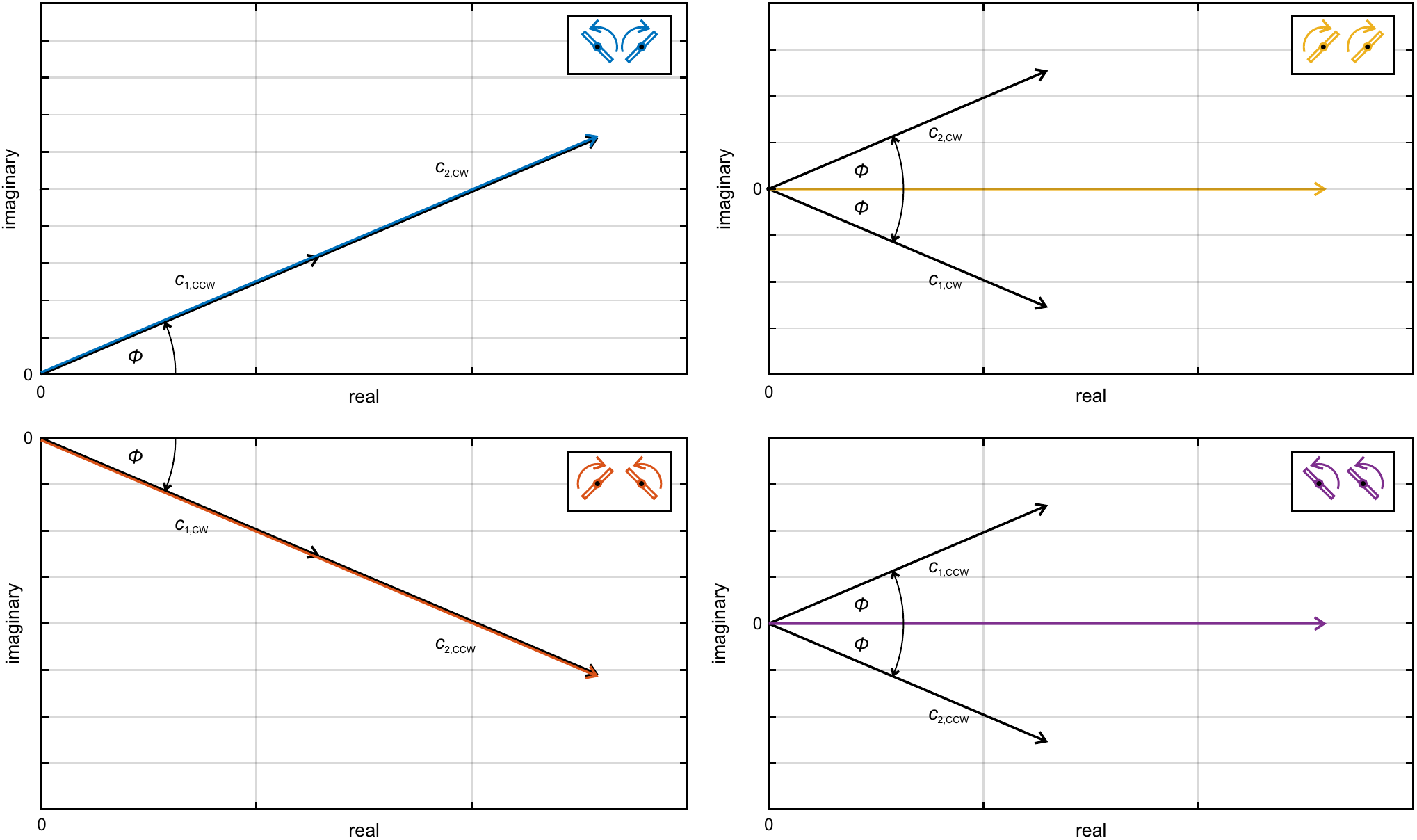} 
    \caption{\textbf{Superposition of modal forces.} Qualitative illustration of the modal forces of each rotating bar (black arrows) in the complex plane for different setup configurations. The total modal force (colored arrows) originates from a complex superposition of the individual modal forces. Phase angle of $\phi=\pm23^\circ$, as for $\Omega=\omega/2$, $d_0 = 300$~mm. }%
    \label{sfig:5}%
\end{figure*}

The situation of the superposition of the force fields generated by the two identical rotating bars has been  described only qualitatively in the main article. To mathematically confirm this description, it is useful to look at the Fourier component at $\omega$ of the individual 1D force components
\begin{eqnarray}
F_{y,b}^{(k)}=\int_0^{l_d}f_{y}^{(k)}U_{b,1}(x_d)\text{d}x_d\approx c_k\ee^{\ii\omega t}\, ,
\end{eqnarray}
that are summed up to build the total modal force $\modalforce$, cf. Eq.~\eqref{eq:one}. The variable $k$ denotes the number of the rotating bar. The Fourier series yields the complex amplitude at the detector's resonance frequency $\omega$, given by the coefficient $c_k$. Assuming identical bars, perfect symmetry with respect to the detector beam and a perfectly symmetric first bending mode shape $U_{b,1}$, the amplitudes $\lvert c_k \rvert$ of both complex forces must be equal. The angle $\angle c_k$, however, differs in sign depending on the rotation direction. It can be shown that, in case of opposite direction, $c_1$ equals $c_2$, hence the total force has twice the amplitude and the same phase as the components $c_k$. Considering the bars rotating in the same direction $c_1=c_2^*$, where $^*$ denotes the complex conjugate. Consequently, $c_1+c_2$ is a real value which means that no phase shift occurs. However, the amplitude gets smaller since the contribution of the imaginary part vanishes. Therefore, a phase angle of $\angle c_k=\pm23^\circ$ yields an amplitude reduction of 8\%, as observed both experimentally and theoretically, cf. Figs.~\ref{fig:2} and \ref{fig:4}. Figure~\ref{sfig:5} illustrates the superposition of the forces qualitatively.\\
The situation of both bars rotating in the same direction is thus very convenient to test both the numerical and experimental results. Numerical errors can be detected easily as a phase shift $\neq 0$, while experimentally measured phase shift variations indicate either crosstalk or a deviation from the ideal situation of two identical, perfectly synchronized, symmetrically oriented bars. Likewise, differences between the bars can be tested by using one rotating bar only. 
\section{Numerical model}
\label{sec:FEM}
The one dimensional Euler-Bernoulli approach assumes a line distribution of the mass of the detector beam. For the setup presented here this approximation gives reasonable results only if the rotating bars are far away from the detector beam ($d_0>0.5$\,m). For improved accuracy, a numerical, 3D finite element (FE) simulation of the transmitter/detector setup is necessary to accurately predict the gravitationally induced motion of the detector beam and to compute the coefficient $\Gamma$ in Eq.~\ref{eq:2} for all distances $d_0$.\\
The detector beam is modeled as a freely suspended, homogeneous, linear-elastic beam, attached to two linear springs in the nodes of its first bending mode. Due to the lock-in measurement technique, the experimentally measured response of the detector is available at the excitation frequency $\omega$ only. Hence, it suffices to solve for the steady state response of the FE model. The damping of the system is modeled using modal damping \cite{hagedorn2007}.\\
Since the force field generated by the rotating bars, illustrated as a transverse force pulse moving along the $x$-axis of the detector beam, represents a moving load problem, it is possible that higher bending modes of the detector are excited as well \cite{Fryba1973}. 
This, in turn, could lead to an additional contribution to the motion measured at the frequency of the first bending mode, which would require an adjustment of Eq.~\ref{eq:2}, which is based on a SDOF approximation of the detector.\\
Therefore, the contribution of higher order modes was investigated numerically: Since the modal damping allows to assign experimentally measured Q factors to the first three in-plane bending modes, that is, $Q_1 = 45000$, $Q_2 = 381$, $Q_3 = 5150$, as well as for the rigid body modes of the beam, $Q_\text{rb} = 2900$, a comparison was performed between a simulation using specific modal Q factors $Q_i$ and a simulation where all modes have the same damping. The results revealed that the influence of higher modes is negligible for the setup presented in this article, thus the first bending mode can be assumed to be fully decoupled from the other modes. Consequently, Eq.~\ref{eq:2} represents a valid approximation of the structural resonance.\\
The coefficient $\Gamma$ has been calculated individually for all different setup configurations and 37 discrete distance values $d_0$ between 290\,mm and 600\,mm using COMSOL Multiphysics 5.6 and MATLAB R2020b. If necessary, amplitude values at the specific distances are interpolated using a power law fit with the coefficients from Table~\ref{tab:2}, while the phase is interpolated linearly.\\
The detector beam is modeled as a linear-elastic rectangular beam with Young's modulus $E=\SI{107.4}{\mega\pascal}$ and Poisson's ratio of $\nu = 0.37$, using the solid mechanics module. The Young's modulus is determined through a fit to the first resonance frequency of the beam. All other properties of the beam are the same as described in~\cite{Brack2022}. The beam is attached to two linear springs in the nodes of its first bending mode. These springs have a nonzero stiffness in $y$-direction only and their stiffness is set such that it matches the experimentally determined resonance frequency of the beam's translational pendulum motion in $y$-direction of $f=858.3$\,mHz.\\
The computation of the detector beam velocity is a three-step process: In a first step, the resonance frequency $\omega_0$ of the beam is computed in the absence of external fields. This resonance frequency is later used to run the frequency space simulation exactly at the resonance frequency of the beam. In a second step, the gravitational force field is computed. The gravitation force on the beam is given by the integral of Newton's force law over the volume of each rotating bar. Since the force is periodic in time, it can be written as a Fourier series, where the Fourier coefficients are computed in every node of the FE model. In the last step, the FE model of the beam is solved in frequency space at the resonance frequency $\omega_0$. Depending on the selection of the Fourier coefficient, the solution corresponds to different rotation speeds $\omega$ of the bars relative to the resonance frequencies. \\ 
A convergence study has been performed to ensure the numerical accuracy of the simulation. The gravitational force of the rotating bars has been computed by discretizing the bars into 396 mass elements. For the ensuing discrete Fourier transform of the force, the rotation period of the bars was evaluated at 64 points per rotation. Finally, the FE model is solved for approximately 35000 tetrahedral elements using quadratic interpolation ($\text{DOF} \approx 170000$). Taking all three discretization steps into account, we have found the numerical error of the simulation to be far below the measurement errors in the experiment and can thus be neglected in the error budget.
\section{\label{Gravitational waves generated by rotating bars}Gravitational waves generated by rotating bars}
The setup of a rotating bar is a classical textbook example of a laboratory quadrupole generator of gravitational waves \cite{misner2017,Giulini2017}, in which the power of a slender bar radiated as gravitational waves can be approximately calculated via
\begin{equation}
L_{\text{GW}} \approx \frac{32}{5}\frac{G}{c^5}\Omega^6I_z^2\,,
\label{eq:power}
\end{equation}
where $I_z$ denotes the rotational inertia of the bar rotating around the $z$ axis with frequency $\Omega$ and the speed of light $c$. For one of the rotating bars presented in this setup rotating with $\Omega=\omega_0/2$, this power calculates to $L_{\text{GW}} \approx 4.2\times 10^{-43}$\,W.\\
In contrast, the power transmitted from two oppositely rotating bars to a detector in a distance $d_0=296$ mm resonating at $\omega_0$ has been estimated to \Ediss, based on the measured resonance amplitude and Q factor of the detector \cite{Brack2022}. \\
Hence it can be concluded that the generated movement of the detector beam is not attributed to gravitational waves, emitted by the rotating beams but by the dynamic gravitational (near) field. 
\section{\label{sec:Rotational System}Rotational system - mechanics and control}
The rotation system is composed of two independent, identical rotary units, of which one system is exemplarily described in this section.\\
The rotating bar is placed on a turntable, where it can be adjusted and fixated using a clamping cover with alignment pins. The bar exhibits a precisely manufactured indentation in its center of rotation for additional fixation. The rotary unit is mounted with two preloaded spindle bearings (HQW Precision GmbH SV7902) that are clamped to a bearing bracket mounted to a massive aluminum base plate of 15\,mm thickness. Attached to the shaft is a maxon EC-i 40 brushless 50\,W electric motor with ENC 16 EASY encoder, connected to the base plate as well. \\
The motor is internally controlled by a maxon EPOS4 50/5 controller. To have full control over the rotation, a master controller based on an Arduino MEGA 2560 is additionally used that communicates with the EPOS system via CAN bus. Both the master and EPOS controller and the power supply are placed outside of the vacuum chamber.\\
For the gravitational excitation it is of utmost importance that the rotation of both bars is synchronized to a reference signal as precisely as possible. For this task, the built-in controller and rotation sensors are not suitable, due to their limited resolution and non-synchronized clock rates. Therefore, two independent light barriers are embedded in the base plate (cf. Fig.~\ref{fig:3}c). The light barriers are built from a high-speed PIN photodiode (SFH 2701, 730\,nm), a 500\,MHz LTC6268 operational amplifier and a 280\,MHz LTC6752 comparator. A TLC555 timer is used as line driver, producing a delay of the pulse of 244\,ns. The emitted light is reflected from a small aluminum patch (0.035\,g) glued to the rotating bar. To ensure mechanical balancing, the same patch is also glued to the opposite side of the bar.\\
To establish a rotation with a frequency $\Omega$ being a fraction of the frequency of a sinusoidal reference signal $\omega$, that is, the frequency of measurement provided by a frequency generator, a comparator converts the reference signal to a rectangular signal of frequency $\Omega$, which can then be fed into a digital, flip-flop-based frequency divider. Finally, a PID controller adjusts the frequency of rotation by synchronizing the edges of both the rectangular signal and the pulse of the light barrier with a time resolution of 62.5\,ns. 
\section{\label{sec:MU}Measurement uncertainty}
Assuming uncorrelated input quantities, the combined standard uncertainty associated with estimating $G$ can be calculated from Eq.~\ref{eq:2} and Table~\ref{tab:2} using a first-order Taylor approximation for each measurement individually. The uncertainties associated with estimating $Q_d$ and $\omega_0$ result directly from the SDOF fit of the measurement. Initially, the distance $d_0$ was manually adjusted, where a systematic error of 0.5\,mm was assumed. The automatic positioning system itself works very precisely with an error of 1\,\textmu m.  As mentioned in \cite{Brack2022}, the uncertainty of the measured velocity $v_{b,0}$ is yet unclear for the extremely small amplitudes relevant in the measurements presented here. Therefore, the uncertainties reported in the data sheets of the laser vibrometers have been used for a first assessment. Since the detector bending amplitude is calculated from a linear combination of three laser measurement signals, the contribution to the combined uncertainty reduces by a factor 0.58. Further, an angular misalignment of the laser beam with respect to the detector/transmitter surface of max. 1$^\circ$ contributes to the uncertainty. Errors that can be attributed to a non-gravitational detector excitation, most likely due to mechanical crosstalk, are estimated with a \crosstalkpercent \% systematic error, based on an evaluation of the detector chamber acceleration. To estimate the variance of the model parameter $\gamma$, a quasi-Monte Carlo method has been applied, using 2000 quasi-random sequences drawn from the probability distributions specified for the input parameters (Sobol method) \cite{Morokoff1995,Sobol1967}. The used input parameters are assumed to be normally distributed with the following standard uncertainties: detector and transmitter dimensions, \beamdimensions; transmitter masses, measured with a Mettler-Toledo XP6002S scale with \masses~combined uncertainty; an uncertainty of $x,\,y$ and $z$ position of the rotation centers of each \rotationCenterOffset; an eccentricity of $\pm$\eccentricity~of the rotating bars, a constant angle offset between the transmitter bars of $\pm$\barangledifference~and an angular misalignment of \anglemisalignment~between the detector's central axis and the transmitter axis given by the centers of rotation. The Monte Carlo simulation has been performed for each setup variation and distance used in the experiments. Therefore, an individual combined uncertainty results for each measurement. In Table~\ref{tab:2} the maximum values are exemplarily reported, resulting in a combined standard error of max. \GSUsingle\ for an individual measurement. \\
In this work, the distance $d_0$ has the highest contribution to the uncertainty of $G$, since the distance uncertainty scales with a factor of up to 5.83, cf. Table~\ref{tab:2}. To achieve an uncertainty $\Delta G/G \ll 1\%$, future activities must therefore focus primarily on the distance $d_0$, crosstalk elimination, the velocity measurement, and the Q factor determination. Further, influences that are not yet included such as material inhomogeneities, form tolerances, numerical errors etc. must be considered as well.
\section{\label{sec:theory_PhaseJitter}Dynamic influence of phase jitter}
To achieve a rotation of both bars exactly at a fraction of the resonance frequency, a control loop is necessary that enables the synchronization of the bars' rotation frequencies to an external reference signal. However, a certain disturbance of said rotation, quantified by a phase jitter $J_p(t)$, might influence the resonance excitation of the detector. Therefore, the influence of a certain phase variation on the detector shall be discussed briefly.\\
To estimate the errors associated with said disturbance, one can reduce the bending motion to a single-degree-of-freedom oscillator with an external, harmonic excitation. This is a valid approach since the Q factor of the first bending mode is very large ($Q_d\approx45000$) and resonance frequencies are well separated. Considering excitation at resonance $\omega_0$, we can introduce the jitter as phase modulation to the excitation, that is,
\begin{equation}
F_{n}=F_0\cos\left(\omega_0t+J_p(t)\right)\,.
\end{equation}
The response $x(t)$ of the oscillator is described by its amplitude and phase $x(t) = A\ee^{\ii\omega_0 t+\ii\phi}$. The effects of the phase noise $J_p(t)$ on the output phase can be analyzed using the transfer function \cite{Rubiola2008,brack2016}
\begin{equation}
\Delta\bar\phi(\omega) = \frac{\omega_0}{2Q_d\ii\omega+\omega_0}\bar{J_p}(\omega) 
\label{eq:phase_amp_dyn}
\end{equation}
where $\Delta\bar\phi$ denotes the absolute variation of the detector's phase and amplitude around the value at resonance in the frequency domain (indicated by the overbar). The response of the amplitude, however, cannot be easily described by the transfer function of a linear time-invariant system. It can be shown, however, that the amplitude approximately shows a similar low-pass behavior \cite{brackdiss2017}.\\
As a consequence, the detector acts as a low-pass filter to disturbances of the excitation phase. Disturbances transmit in the phase via a first order low-pass filter with a cut-off frequency of $\omega_c=\omega_0/(2Q_d)\approx0.5$\,mHz. Numerical simulations with MATLAB Simulink show a relative amplitude deviation $<1\times10^{-4}$ for a white phase noise corresponding to 0.03$^\circ$ RMS jitter.
In summary, phase variations of the excitation force have a negligible effect on the detector's behavior, since they are minimized by the inertia of the transmitter bars, the motor control system and the high-Q detector itself. 
\begin{figure*}[h!]
    \centering
				  \begin{overpic}[width=0.9\textwidth]{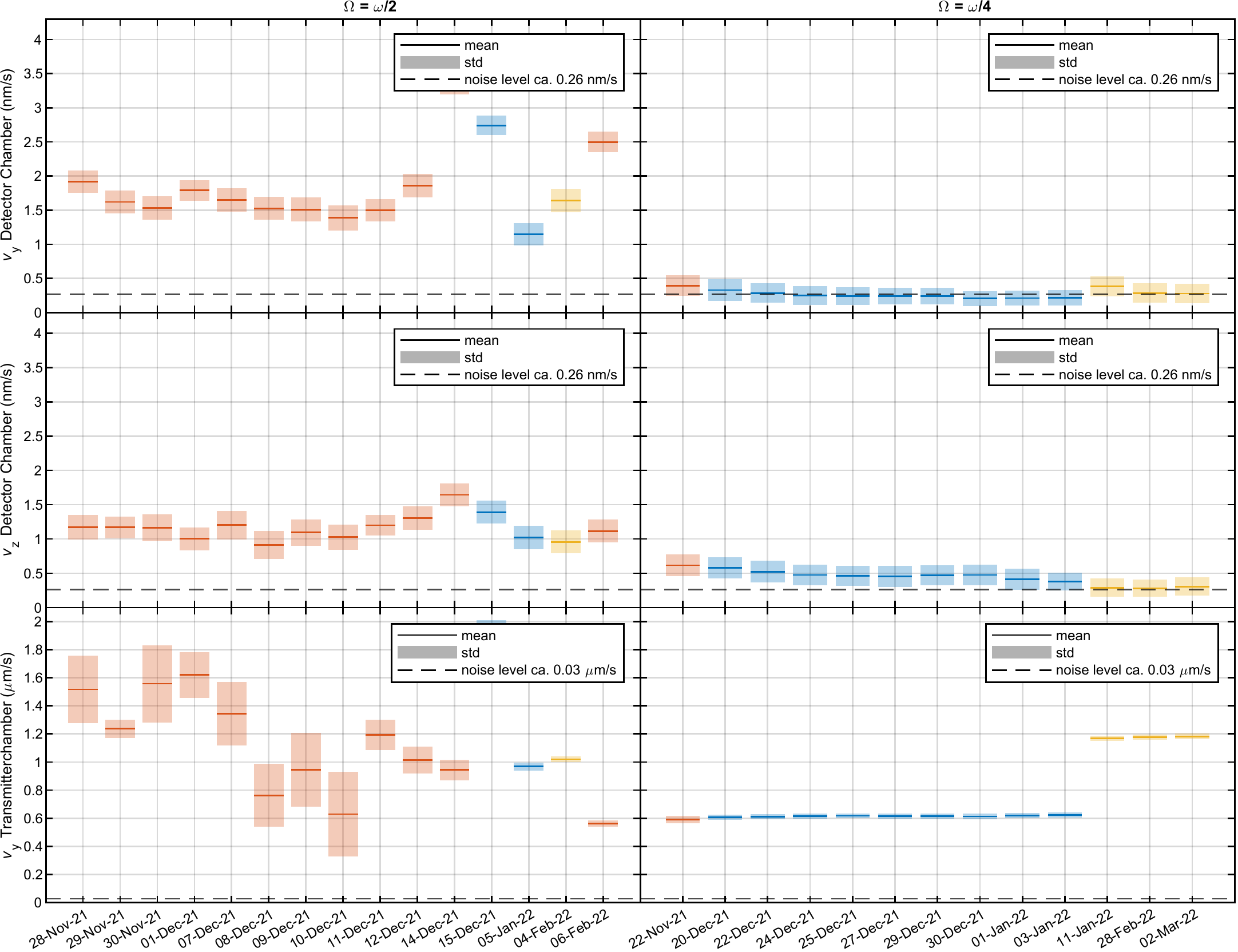}
   
		\put(69.35,21){\includegraphics[scale=0.5]{legend}}
\put(69.35,44.5){\includegraphics[scale=0.5]{legend}}
\put(69.35,68){\includegraphics[scale=0.5]{legend}}
		\put(21,21){\includegraphics[scale=0.5]{legend}}
\put(21,44.5){\includegraphics[scale=0.5]{legend}}
\put(21,68){\includegraphics[scale=0.5]{legend}}
  \end{overpic}
    \caption{\textbf{Detector and transmitter chamber movement}. Average velocity amplitude at the measurement frequency $\omega$ measured with lock-in amplifiers (8th order lowpass, 31 s time constant, 16 min averaging) \textbf{First two rows}, Velocity amplitudes of the detector chamber both in $y$ and $z$ direction, measured with a Kinemetrics EPI ES-T FBA triaxial accelerometer. \textbf{Third row}, Velocity amplitude in $y$ direction of the transmitter chamber, measured with a Bruel\&Kjær 4535-B-001 triaxial accelerometer.  The dashed lines represent the noise level of the measurement. \textbf{Left column}, Excitation with $\Omega=\omega/2$. \textbf{Right column}, Excitation with $\Omega=\omega/4$.}%
    \label{sfig:1}%
\end{figure*}
\begin{figure*}[h!]
    \centering
				  \begin{overpic}[width=0.9\textwidth]{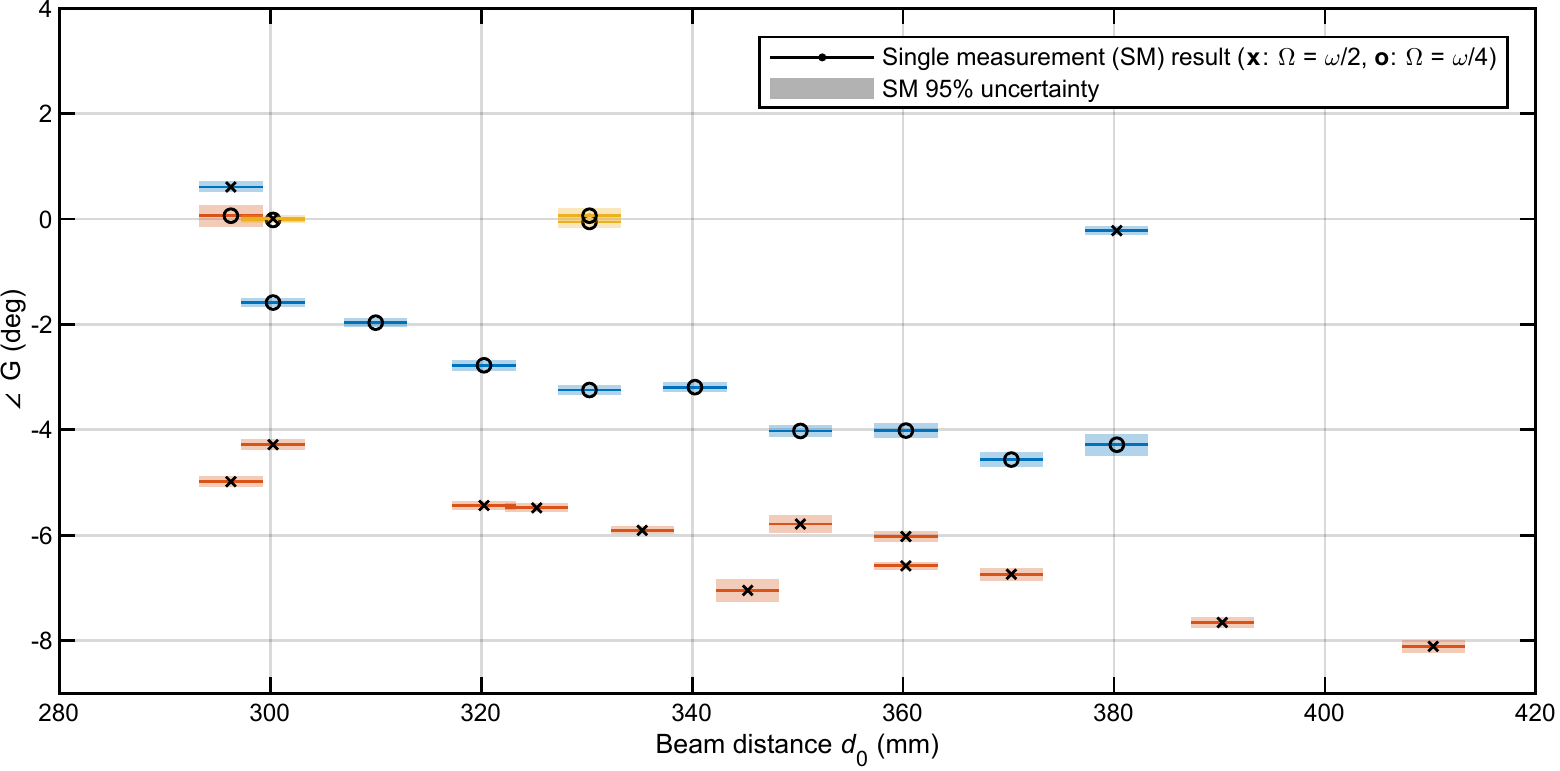}
					\put(87.6,33){\includegraphics[scale=0.5]{legend}}
  \end{overpic}
    \caption{\textbf{Phase difference between measurement and theory} as a function of transmitter/detector distance, $d_0$. From Eq.~\ref{eq:2}, the phase difference between theory and measurement can be determined as the angle of the resulting complex value of $G$. Both the theoretical and experimental phase have been expressed relative to a reference phase, that is, the phase at same rotation direction (yellow colour coded results). The experimental reference phase has been determined by an average of the corresponding results, separately for each harmonic. The y-extent of the error patches denote statistical uncertainties of the phase, the x-range has no numerical meaning but is for illustrative purposes only. The illustration indicates an increasing phase difference for increasing distance, $d_0$, both for the second (\textbf{x}) and fourth (\textbf{o}) harmonic. A certain, however yet unknown, systematic error seems to be present, which is most likely due to remaining crosstalk (mechanical, acoustical). For the setup configuration CW-CCW (red color), the error seems to be a bit higher.}%
    \label{sfig:2}%
\end{figure*}
\clearpage
\begin{sidewaystable}[]
\small
\begin{tabular}{@{}llllllll@{}}
\toprule
\textbf{Parameter}                              & \textbf{Unit} & \multicolumn{2}{c}{\textbf{Transmitter 1}}   &\multicolumn{2}{c}{\textbf{Transmitter 2}} & \multicolumn{2}{c}{\textbf{Detector}}     \\ \midrule
\textbf{}                                       & \textbf{}     & \textbf{Value} & \textbf{SU}  & \textbf{Value} & \textbf{SU} & \textbf{Value} & \textbf{SU} \\
Mass                                            & g             & 970.8         & 0.1       & 971.8         & 0.1                    & 647.7         & 0.1                          \\
Length                                          & mm            & 499.86        & 0.01        & 499.86        & 0.01                  & 1000.00           & 0.01                         \\
Width                                           & mm            & 10.06          & 0.01          & 10.06          & 0.01                & 8.49           & 0.01                          \\
Height                                          & mm            & 10.06          & 0.01           & 10.07          & 0.01               & 16.97          & 0.01                          \\
Cross sectional area*                           & mm$^2$          & 101.2         & 0.14           & 101.3          & 0.14                 & 144.1          & 0.19                           \\
Second moment of inertia \\ with respect to z-axis*  & $10^{-10}$m$^4$           &          &             &         &                & 8.65       & 0.03                         \\ 
Resonance frequency @ 11.4$^\circ$C                    & Hz            &           &                 &           &                    & 42.650759      & 0.000015                      \\
Mass per unit length*                        & kg/m          &          &            &          &             & 0.6477         & 0.0001                       \\
Bending stiffness*                           & Pa\,m$^4$       &           &                &        &            & 92.924         & 0.014                         \\
Young's modulus*                               & GPa           &             &             &             &                  & 107.4          & 0.4                           \\
\midrule
Transmitter distance $x_0$ & mm& 800              &       0.3      &             &                  &           &                          \\
\bottomrule
\end{tabular}

\caption{Detector and transmitter model parameters and corresponding standard uncertainties (SU).\\
$^*$Derived parameters.}
\label{tab:extab1}
\end{sidewaystable}
\clearpage
\bibliography{bibliography}

\end{document}